# Atomic-resolution structural and spectroscopic evidence for the synthetic realization of two-dimensional copper boride

Hui Li[1], Qiyuan Ruan[2], Cataldo Lamarca[1], Albert Tsui[1], Boris I. Yakobson[2,*] and Mark C. Hersam[1,3,4,*]

[1]*Department of Materials Science and Engineering, Northwestern University, Evanston, IL 60208, USA*
[2]*Department of Materials Science and NanoEngineering, Rice University, Houston, TX 77005, USA*
[3]*Department of Chemistry, Northwestern University, Evanston, IL 60208, USA*
[4]*Department of Electrical and Computer Engineering, Northwestern University, Evanston, IL 60208, USA*

*Corresponding authors: biy@rice.edu and m-hersam@northwestern.edu

**Abstract**

Since the first realization of borophene on Ag(111), two-dimensional (2D) boron nanomaterials have attracted significant interest due to their polymorphic diversity and potential for hosting solid-state quantum phenomena. Here, we use atomic-resolution scanning tunneling microscopy (STM) and field-emission resonance (FER) spectroscopy to elucidate the structure and properties of atomically thin boron phases grown on Cu(111). Specifically, FER spectroscopy reveals unique charge transfer and electronic states compared to the distinct borophene phases observed on silver, suggesting that the deposition of boron on copper can result in strong covalent bonding characteristic of a 2D copper boride. This conclusion is reinforced by detailed STM characterization of line defects that are consistent with density functional theory (DFT) calculations for atomically thin $Cu_8B_{14}$. This evidence for 2D copper boride is likely to motivate future synthetic efforts aimed at expanding the relatively unexplored family of atomically thin metal boride materials.

**INTRODUCTION**

The pioneering realization of borophene polymorphs on silver (*1, 2*) demonstrated the feasibility of growing boron nanomaterials in two-dimensional (2D) forms, inspiring additional efforts to achieve borophene synthesis on other substrates including Al (*3*), Au (*4*), Ir (*5-7*), Cu (*7-11*), and Ru (*12*). A key element in stabilizing borophene is charge transfer with the substrate (*13*), where metals can act as a charge reservoir to compensate for the fact that elemental boron materials struggle to satisfy the octet rule. While many efforts to diversify the set of growth substrates have resulted in distinct borophene phases, another possibility is that boron will covalently bond with the underlying metal to form atomically thin metal borides. Distinguishing between borophene and 2D metal borides is often challenging, necessitating the utilization of multiple atomic-resolution structural and spectroscopic characterization methods.

Of particular interest is the product of atomically thin boron deposition on Cu(111), where different interpretations of the observed structure have been reported including single-layer borophene (*7-9*), bilayer borophene (*11*), and 2D copper boride (*14-16*). Beyond the well-known superconductor $MgB_2$ (*17, 18*), crystalline metal borides with layered structures are rare, which implies that 2D metal borides are relatively unexplored compared to other 2D materials despite theoretical predictions that they should have superlative electronic, magnetic, and mechanical properties (*19, 20*). Therefore, it is of high interest to show strong evidence for the synthetic realization of 2D copper boride to motivate future efforts aimed at expanding the relatively unexplored family of atomically thin metal boride materials.

Here, we employ atomic-resolution scanning tunneling microscopy (STM) and field-emission resonance (FER) spectroscopy to characterize the structure and properties of 2D boron phases grown on Cu(111). FER spectroscopy reveals charge transfer and electronic states that are consistent with strong covalent bonding between boron and copper that is characteristic of a 2D copper boride. Moreover, STM characterization of line defects agrees with density functional theory (DFT) calculations for atomically thin $Cu_8B_{14}$ as opposed to a distinct borophene adlayer. In this manner, this work not only clarifies the structure of atomically thin boron deposited on Cu(111) but also provides a methodology for differentiating borophene from atomically thin borides on other metal substrates.

**RESULTS**

**Field-emission resonance spectroscopy of atomically thin boron on Cu(111)**

Atomically thin boron phases on Cu(111) single crystals are achieved by electron-beam deposition from a high-purity boron rod onto an atomically clean Cu(111) surface maintained at a substrate temperature of ~550 °C in ultra-high vacuum (UHV). Fig. 1A shows a representative STM image of the Cu(111) surface after atomically thin boron deposition, in which a spatial derivative along the horizontal direction is taken to enhance imaging contrast. A highly-ordered 2D sheet is observed with a morphology consistent with earlier reports as mentioned above. The height profile extracted from a zoomed-in area (fig. S1) compares the thickness of the 2D sheet with a pristine copper step height, confirming sub-monolayer coverage. With a sharper tip, an atomic-resolution STM image is obtained. As seen in Fig. 1B, the topography possesses a zigzag superstructure with ~90° turns. The higher brightness at the corners can be explained by the higher local density of states (LDOS) resulting from the compressed atomic spacing at the ~90° turning points. Due to convolution between electronic and physical structure, the STM image alone is insufficient to identify the identity of this zigzag phase. Consequently, we will refer to the as-grown structure as B/Cu.

Since a well-defined boundary exists between pristine Cu(111) and the B/Cu domain, characterization of the electronic properties with spatially resolved spectroscopic measurements in this area is desirable as it can ensure that information is extracted under the same tip condition. In particular, we employ field-emission resonance (FER) spectroscopy to extract spatially resolved electronic properties. FER spectroscopy is realized using the constant-current mode of scanning tunneling spectroscopy (STS), where both $\partial I/\partial V$ and $\partial z/\partial V$ signals provide a series of clear resonance peaks. Since its early demonstrations (*21, 22*), FER spectroscopy has proven to be a highly reliable and insightful method for the study of local work function variations of engineered metal surfaces (*23-28*). In our case, we performed FER spectroscopic mapping (see movie S1 for the full video) over the field of view in Fig. 1C where a $\partial I/\partial V$ spectrum up to 10.0 V is registered on each point of a grid map using a lock-in amplifier. A slice of the map taken at 4.5 V is shown in Fig. 1D that clearly highlights the FER intensity difference between the Cu(111) and B/Cu regions. These systematic distinctions can be further observed from a series of spectra (Fig. 1E) extracted along the black line in Fig. 1C and Fig. 1D, where the purple and gray lines trace the evolution of the first and second order Rydberg peaks, respectively. The peaks that correspond to the image potential states exhibit an unambiguous shift to lower energies from the pure Cu(111) area to the B/Cu area. Since image potential states are positively correlated with the local work function (LWF) (*29*), the FER spectra enable a direct visualization of LWF shifts. A quantitative extraction of the LWF can be achieved by fitting to the following equation based on resonant tunneling in the Fowler-Nordheim regime (*30*):

$$eV_n = \phi + (\frac{3\pi\hbar eF}{2\sqrt{2m}})^{2/3} n^{2/3} \tag{1}$$

where $V_n$ is the energy of the peak of order $n$, and the effective electric field ($F$) is approximated as a constant. The LWF ($\phi$) is then extracted by fitting the $n > 2$ peaks with respect to $n^{2/3}$. This fitting results in locally averaged work functions of $\phi_{Cu} \sim 4.82$ V and $\phi_{B/Cu} \sim 4.45$ V. The Cu(111) LWF is in agreement with previous literature (*31*), thus validating the tip condition. The smaller LWF of B/Cu compared to Cu(111) is consistent with the peak position shifts in Fig. 1E. It is, however, contrary to the cases of most distinct borophene phases on metal surfaces, in which the charge transfer from the metal substrate to the borophene layer results in a surface dipole with an increased LWF (*28, 32*).

Another notable feature in the FER spectra is the presence of additional peaks below 4.0 V that only exist in the B/Cu region. In the FER spectrum of the pristine Cu(111) region (Fig. 1F), the $n = 1$ peak is found at ~4.5 V with no additional features other than resonance peaks, which is consistent with previous studies on the Cu(111) surface (*23, 25*). To reveal the nature of the low energy peaks in the B/Cu FER spectra, a measurement as a function of tip-sample distance was performed by varying the tunneling junction resistance at a fixed point. As shown in Fig. 1G, the higher-order FER peaks shift when the setpoint current changes (setpoint bias voltage held fixed at 100 mV), which is expected for resonance states (*28*). On the other hand, the peak positions near 1.96 V and 3.35 V are almost constant, suggesting that these electronic states are providing additional information about the B/Cu region beyond LWF. In principle, constant-height $\partial I/\partial V$ measurements should also be able to resolve these electronic states as demonstrated in fig. S2. However, the overall high LDOS makes it difficult to precisely extract the constant-height $\partial I/\partial V$ peaks. Therefore, FER spectroscopy is the superior approach in this context. The nearly invariant low-energy FER spectral features for B/Cu are similar to copper nitride films with varied stoichiometries (*33-35*). In addition, constant-current $\partial I/\partial V$ measurements at negative energies for B/Cu (fig. S3) exhibit similar peaks compared to $Cu_3N$ (*35*). These spectral similarities between B/Cu and covalently bonded copper nitrides suggests that the B/Cu region is likely a covalently bonded copper boride.

**DFT calculations and simulations**

To investigate the B/Cu system from a theoretical perspective, DFT calculations and simulations were performed on multiple plausible B/Cu structures including a distinct single-layer borophene phase and a covalently bonded $Cu_8B_{14}$ copper boride phase. In particular, $v_{1/6}$-borophene ($\beta_{12}$) and $Cu_8B_{14}$ copper boride on Cu(111) are shown in Fig. 2A and Fig. 2B, respectively. The $v_{1/6}$ borophene layer has a rectangular unit cell, whereas the $Cu_8B_{14}$ consists of a superlattice ($a \sim 2.18$ nm, $b \sim 1.60$ nm) depicted by a red rhomboid in Fig. 2B, in which the corner boron atoms sit atop the second layer copper atoms. From the side view of $Cu_8B_{14}$, it can be seen that the B atoms are imbedded in the first layer Cu atoms and inhomogeneous strain exists due to the lattice mismatch. Fig. 2C and Fig. 2D depict the DFT-calculated charge redistributions in these two scenarios. Clear differences in charge transfer are apparent between $v_{1/6}$ borophene and $Cu_8B_{14}$, with the $Cu_8B_{14}$ case resulting in a decreased LWF that is consistent with the aforementioned experimental observation. Simulated STM images of $v_{1/6}$ borophene and $Cu_8B_{14}$ are also provided in Fig. 2E and Fig. 2F, respectively. The simulated STM image of $Cu_8B_{14}$ shows a periodic zigzag pattern that closely matches the experimental STM image. Fig. 2G further highlights the agreement between experiment and theory for $Cu_8B_{14}$ by overlapping the atomically-resolved experimental STM image (reproduced from Fig. 1B) with the DFT-calculated $Cu_8B_{14}$ structure.

**Line defects for atomically thin boron on Cu(111)**

Previous reports for atomically thin boron deposition on copper have focused on the pristine regions of the 2D layers without explicitly addressing defect structures, despite the value of defect structures in identifying the atomic structure of surface adlayers (*36, 37*). Consequently, we located line defects in B/Cu and characterized their topographic and electronic properties at the atomic scale. Fig. 3A shows two parallel line defects on a B/Cu terrace. A high-resolution STM topographic image (inset) of the white dashed box reveals the atomic-scale structure of the line defect, where the ~90° turns take place at the ends of longer atomically straight chains compared to the shorter atomically straight chains in the undefective regions. Fig. 3B provides another STM image across the double line defects taken with a sharper STM tip, where the atomic chains appear as valley-bridge intermixing patterns due to the participation of different electronic states in the tunneling junction. These observations do not resemble previously reported line defects for borophene (*34*) and are likely the result of the mismatch with the subsurface copper or the local disorder of the B/Cu layer.

To further assess the electronic properties of the line defects in B/Cu, a constant-current $\partial I/\partial V$ map at 3.3 V is shown in Fig. 3C (see movie S2 for the full video). A prominent observation is the emergence of the clear rhomboid lattice (marked in red) in the undefective regions. The lattice constants ($a \sim 2.14$ nm, $b \sim 1.57$ nm) match the DFT-calculated $Cu_8B_{14}$ superlattice discussed above. The $\partial I/\partial V$ maps at other energies show different patterns, but with similar periodicities (fig. S4). In contrast, the line defects exhibit distinct $\partial I/\partial V$ intensities and features, which is more clearly shown in Fig. 3D, where the $\partial I/\partial V$ spectra of five representative locations (colored circles in Fig. 3C) are plotted together for direct comparison. The unique characteristics of the line defects include the peaks near 1.96 V (4.15 V) shifting to higher (lower) energy and the disappearance of the peak near 3.3 V that was present in the undefective regions. For better visualization, two series of $\partial I/\partial V$ spectra (from 1.0 V to 5.0 V) along the black and gray lines are shown in Fig. 3E and Fig. 3F, respectively. In these spectra, the overall anti-phase relationships are evident. Specifically, when the peak near 3.3 V is stronger, the intensities of the 1.96 V and 4.15 V peaks are correspondingly lower. The unique electronic characteristics of the line defects are especially clear in Fig. 3F, where the stronger intensities of the 1.96 V and 4.15 V peaks dwarf the peak near 3.3 V.

**Energy maps and correlations of electronic states**

In addition to the peak intensities, the peak positions also contain important information about the electronic states. With a precise spatial registration, the energy positions of these peaks on each point can be extracted from the spectroscopic mapping. Figs. 4A-4C show the energy maps of those observed peaks below 5.0 V. The line defects appear as long stripes in the 1.96 V and 4.15 V peak maps, while the kink positions in the spectra are selected for the extraction of the 3.35 V peak map due to the absence of obvious intensity peaks. Moreover, by applying the fitting with Eq. (1), the local work function of each point can also be extracted to construct the LWF map, as shown in Fig. 4D. An overall lower LWF is again observed from the line defects, suggesting different charge transfer characteristics for the longer atomically straight chains. Spatial modulations with lattice resolution are present in all of these maps, which further confirm the correlations of the observed peaks (with detailed cross-correlation analysis shown in fig. S5), indicating the likely competition between different electronic states.

As discussed above, the peak near 3.3 V almost disappears at the line defects. The absence of such peaks on defects can be expected when the LDOS of the boron $p_z$ orbitals shifts to higher energies (to merge with the $n = 1$ image potential state) due to a larger amount of charge transfer from copper, which also qualitatively agrees with the LWF results. The effects of atomic orbitals on the correlations of peaks in constant-current spectra have been discussed in previous reports (*33-35*). Specifically, the anti-phase relationships of similar peaks in $Cu_3N$ films were attributed to the half-unit cell shift between different N atoms (*35*). Due to the similar spectral behavior for $Cu_3N$ and B/Cu, it is likely that the peak near 3.3 V for B/Cu can be attributed to the boron $p_z$ orbital state while the 1.96 V peak is related to an interface state. The anti-correlations between the low energy peaks and the higher energy image potential state can then be expected accordingly, where differences in the charge distributions of different boron orbitals due to local changes in the chemical environment near defects determines the appearance of corresponding spectral features.

The different topographic and spectral characteristics in the line defect regions suggest a distinct assembly of boron and copper atoms in the atomically straight chains. Based on the STM images, we constructed a 2D structure including long atomically straight chains and adjacent shorter atomically straight chains with ~90° turns to model the line defect. The left panel of Fig. 4E shows the structural model with a unit cell outlined with a black rectangle. The middle and right panels are the high-resolution experimental STM image and simulated STM image, respectively. The agreement between experiment and theory demonstrates that the proposed structure captures the characteristics of the line defects. Moreover, a $Cu_{11}B_{18}$ stoichiometry in the line defect unit cell suggests a higher concentration of copper compared to pristine $Cu_8B_{14}$, which also aligns with the charge transfer mechanism that underlies the disappearance of the peaks near 3.3 V. For the 1.96 V peak, the shifts in the LDOS of the boron $p_z$ orbital competes with the electron occupations in the $p_x$ and $p_y$ orbitals, thus also altering the interface states.

**DISCUSSION**

In conclusion, we have used UHV electron-beam deposition of boron on Cu(111) single crystals to achieve highly-ordered 2D B/Cu sheets. Using spatially resolved FER spectroscopic mapping, unique electronic states and LWF variations are identified at the atomic scale. The smaller LWF of B/Cu compared to pristine Cu(111) deviates from the typical behavior of a distinct borophene layer on metallic substrates. In addition, the emergence of multiple low-energy FER peaks suggests strong bonding between boron and copper atoms as expected for a 2D copper boride phase. Comparison of these observations with DFT calculations strongly suggests that the 2D copper boride phase is $Cu_8B_{14}$. Further corroborating evidence is obtained by closely analyzing the physical and electronic structure of line defects for B/Cu. In particular, the line defects are inconsistent with previous observations for distinct borophene layers on metal substrates and are instead attributed to local copper-rich regions in $Cu_8B_{14}$ as confirmed by DFT-simulated STM images. Overall, this atomically resolved methodology provides strong evidence for the experimental realization of 2D copper boride. By providing a protocol for differentiating between borophene and 2D metal borides, this work can guide synthetic efforts aimed at identifying other 2D metal borides, thus accelerating research for a relatively underexplored family of 2D materials.

## MATERIALS AND METHODS

### Sample growth
The Cu(111) single crystal was cleaned by repeated Ar$^+$ sputtering (1 kV, $5.0 \times 10^{-5}$ mbar) and annealing (~550 °C) in a UHV chamber (base pressure $< 2.0 \times 10^{-10}$ mbar). Boron deposition was then performed by electron-beam evaporation of a pure boron rod (ESPI metals, 99.9999% purity) on the heated substrate (~550 °C). The accelerating voltage for the evaporator (FOCUS EFM3) was 1800 V, and the filament current was 1.62 A. With a ~5 min boron deposition under a flux ~ 9.2 nA, a 60-70% coverage of B/Cu domains was obtained on the Cu(111) surface.

### STM/STS measurements
All STM/STS measurements were performed with a Scienta Omicron LT STM at ~4 K. The PtIr tip was conditioned by repeated scanning and tip conditioning on the clean Cu(111) surface. The feedback loop was engaged during FER measurements and disengaged for $\partial I/\partial V$ measurements, where both cases were recorded with a SR850 lock-in amplifier with 5 mV bias modulation and 820 Hz frequency. The mapping was conducted only when repeated topographic scans in the interested field-of-view confirmed that sample drift was negligible. Gaussian fittings were used to extract the peak positions, and LWF was extracted by fitting the image potential states with respect to $n^{2/3}$. Cross-correlation analyses were conducted by the scipy.signal.correlate2d package.

### DFT calculations
Structural optimizations and charge distribution calculations were performed by the Vienna Ab initio Simulation Package (VASP) (*38*), adopting generalized gradient approximation with the Perdew–Burke–Ernzerhof (PBE) (*39*) exchange-correlation functional in the framework of the projector augmented wave (PAW) method (*40*). The kinetic energy cutoff was set at 520 eV. The positions of atoms were relaxed until the force of reach atom reached 0.01 eV/Å. The k-mesh was set as 9×5×1 for $v_{1/6}$ borophene and 3×3×1 for $Cu_8B_{14}$ copper boride. A vacuum layer of 15 Å was chosen perpendicular to slab surfaces to avoid interactions from the periodic boundary conditions.

REFERNECES

1. A. J. Mannix, X.-F. Zhou, B. Kiraly, J. D. Wood, D. Alducin, B. D. Myers, X. Liu, B. L. Fisher, U. Santiago, J. R. Guest, M. J. Yacaman, A. Ponce, A. R. Oganov, M. C. Hersam, N. P. Guisinger, Synthesis of borophenes: Anisotropic, two-dimensional boron polymorphs. *Science* **350**, 1513–1516 (2015).
2. B. Feng, J. Zhang, Q. Zhong, W. Li, S. Li, H. Li, P. Cheng, S. Meng, L. Chen, K. Wu, Experimental realization of two-dimensional boron sheets. *Nat. Chem.* **8**, 563–568 (2016).
3. W. Li, L. Kong, C. Chen, J. Gou, S. Sheng, W. Zhang, H. Li, L. Chen, P. Cheng, K. Wu, Experimental realization of honeycomb borophene. *Sci. Bull.* **63**, 282–286 (2018).
4. B. Kiraly, X. Liu, L. Wang, Z. Zhang, A. J. Mannix, B. L. Fisher, B. I. Yakobson, M. C. Hersam, N. P. Guisinger, Borophene synthesis on Au(111). *ACS Nano* **13**, 3816–3822 (2019).
5. N. A. Vinogradov, A. Lyalin, T. Taketsugu, A. S. Vinogradov, A. Preobrajenski, Single-phase borophene on Ir(111): formation, structure, and decoupling from the support. *ACS Nano* **13**, 14511–14518 (2019).
6. K. M. Omambac, M. Petrović, P. Bampoulis, C. Brand, M. A. Kriegel, P. Dreher, D. Janoschka, U. Hagemann, N. Hartmann, P. Valerius, T. Michely, F. J. Meyer zu Heringdorf, M. Horn-von Hoegen, Segregation-enhanced epitaxy of borophene on Ir(111) by thermal decomposition of borazine. *ACS Nano* **15**, 7421–7429 (2021).
7. M. G. Cuxart, K. Seufert, V. Chesnyak, W. A. Waqas, A. Robert, M.-L. Bocquet, G. S. Duesberg, H. Sachdev, W. Auwärter, Borophenes made easy. *Sci. Adv.* **7**, eabk1490 (2021).
8. R. Wu, I. K. Drozdov, S. Eltinge, P. Zahl, S. Ismail-Beigi, I. Božović, A. Gozar, Large-area single-crystal sheets of borophene on Cu(111) surfaces. *Nat. Nanotechnol.* **14**, 44–49 (2019).
9. R. Wu, A. Gozar, I. Božović, Large-area borophene sheets on sacrificial Cu(111) films promoted by recrystallization from subsurface boron. *npj Quantum Mater.* **4**, 1–6 (2019).
10. R. Wu, S. Eltinge, I. K. Drozdov, A. Gozar, P. Zahl, J. T. Sadowski, S. Ismail-Beigi, I. Božović, Micrometre-scale single-crystalline borophene on a square-lattice Cu(100) surface. *Nat. Chem.* **14**, 377–383 (2022).
11. C. Chen, H. Lv, P. Zhang, Z. Zhuo, Y. Wang, C. Ma, W. Li, X. Wang, B. Feng, P. Cheng, X. Wu, K. Wu, L. Chen, Synthesis of bilayer borophene. *Nat. Chem.* **14**, 25–31 (2022).
12. P. Sutter, E. Sutter, Large-scale layer-by-layer synthesis of borophene on Ru(0001). *Chem. Mater.* **33**, 8838–8843 (2021).
13. Y. V. Kaneti, D. P. Benu, X. Xu, B. Yuliarto, Y. Yamauchi, D. Golberg, Borophene: Two-dimensional boron monolayer: synthesis, properties, and potential applications. *Chem. Rev.* **122**, 1000–1051 (2022).
14. X.-J. Weng, X.-L. He, J.-Y. Hou, C.-M. Hao, X. Dong, G. Gao, Y. Tian, B. Xu, X.-F. Zhou, First-principles prediction of two-dimensional copper borides. *Phys. Rev. Mater.* **4**, 074010 (2020).
15. C. Yue, X.-J. Weng, G. Gao, A. R. Oganov, X. Dong, X. Shao, X. Wang, J. Sun, B. Xu, H.-T. Wang, X.-F. Zhou, Y. Tian, Formation of copper boride on Cu(111). *Fundamental Research* **1**, 482–487 (2021).
16. X.-J. Weng, J. Bai, J. Hou, Y. Zhu, L. Wang, P. Li, A. Nie, B. Xu, X.-F. Zhou, Y. Tian, Experimental evidence of surface copper boride. *Nano Res.* **16**, 9602−9607 (2023).
17. C. Buzea, T. Yamashita, Review of the superconducting properties of $MgB_2$. *Supercond. Sci. Technol.* **14**, R115 (2001).
18. H. J. Choi, D. Roundy, H. Sun, M. L. Cohen, S. G. Louie, The origin of the anomalous superconducting properties of $MgB_2$. *Nature* **418**, 758–760 (2002).

19. S. Carenco, D. Portehault, C. Boissière, N. Mézailles, C. Sanchez, Nanoscaled metal borides and phosphides: recent developments and perspectives. *Chem. Rev.* **113**, 7981–8065 (2013).
20. P. R. Jothi, K. Yubuta, B. P. T. Fokwa, A simple, general synthetic route toward nanoscale transition metal borides. *Adv. Mater.* **30**, 1704181 (2018).
21. G. Binnig, K. H. Frank, H. Fuchs, N. Garcia, B. Reihl, H. Rohrer, F. Salvan, A. R. Williams, Tunneling spectroscopy and inverse photoemission: image and field states. *Phys. Rev. Lett.* **55**, 991–994 (1985).
22. R. S. Becker, J. A. Golovchenko, B. S. Swartzentruber, Electron interferometry at crystal surfaces. *Phys. Rev. Lett.* **55**, 987–990 (1985).
23. D. B. Dougherty, P. Maksymovych, J. Lee, J. T. Yates, Local spectroscopy of image-potential-derived states: from single molecules to monolayers of benzene on Cu(111). *Phys. Rev. Lett.* **97**, 236806 (2006).
24. H.-C. Ploigt, C. Brun, M. Pivetta, F. Patthey, W.-D. Schneider, Local work function changes determined by field emission resonances: NaCl/Ag(100). *Phys. Rev. B* **76**, 195404 (2007).
25. S. Joshi, D. Ecija, R. Koitz, M. Iannuzzi, A. P. Seitsonen, J. Hutter, H. Sachdev, S. Vijayaraghavan, F. Bischoff, K. Seufert, J. V. Barth, W. Auwärter, Boron nitride on Cu(111): an electronically corrugated monolayer. *Nano Lett.* **12**, 5821–5828 (2012).
26. F. Schulz, R. Drost, S. K. Hämäläinen, T. Demonchaux, A. P. Seitsonen, P. Liljeroth, Epitaxial hexagonal boron nitride on Ir(111): a work function template. *Phys. Rev. B* **89**, 235429 (2014).
27. M. Schwarz, A. Riss, M. Garnica, J. Ducke, P. S. Deimel, D. A. Duncan, P. K. Thakur, T.-L. Lee, A. P. Seitsonen, J. V. Barth, F. Allegretti, W. Auwärter, Corrugation in the weakly interacting hexagonal-BN/Cu(111) system: structure determination by combining noncontact atomic force microscopy and X-ray standing waves. *ACS Nano* **11**, 9151–9161 (2017).
28. X. Liu, L. Wang, B. I. Yakobson, M. C. Hersam, Nanoscale probing of image-potential states and electron transfer doping in borophene polymorphs. *Nano Lett.* **21**, 1169–1174 (2021).
29. K. Wandelt, The local work function: concept and implications. *Appl. Surf. Sci.* **111**, 1–10 (1997).
30. O. Yu. Kolesnychenko, Yu. A. Kolesnichenko, O. I. Shklyarevskii, H. van Kempen, Field-emission resonance measurements with mechanically controlled break junctions. *Physica B: Condensed Matter* **291**, 246–255 (2000).
31. H. B. Michaelson, The work function of the elements and its periodicity. *J. Appl. Phys.* **48**, 4729–4733 (1977).
32. X. Liu, Q. Li, Q. Ruan, M. S. Rahn, B. I. Yakobson, M. C. Hersam, Borophene synthesis beyond the single-atomic-layer limit. *Nat. Mater.* **21**, 35–40 (2022).
33. C. D. Ruggiero, T. Choi, J. A. Gupta, Tunneling spectroscopy of ultrathin insulating films: CuN on Cu(100). *Appl. Phys. Lett.* **91**, 253106 (2007).
34. T. Choi, C. D. Ruggiero, J. A. Gupta, Tunneling spectroscopy of ultrathin insulating $Cu_2N$ films, and single Co adatoms. *J. Vac. Sci. Technol. B* **27**, 887–890 (2009).
35. K. Bhattacharjee, X.-D. Ma, Y. Q. Zhang, M. Przybylski, J. Kirschner, Electronic structure of the corrugated $Cu_3N$ network on Cu(110): Tunneling spectroscopy investigations. *Surf. Sci.* **606**, 652–658 (2012).
36. A. Eckmann, A. Felten, A. Mishchenko, L. Britnell, R. Krupke, K. S. Novoselov, C. Casiraghi, Probing the nature of defects in graphene by Raman spectroscopy. *Nano Lett.* **12**, 3925–3930 (2012).
37. X. Liu, Z. Zhang, L. Wang, B. I. Yakobson, M. C. Hersam, Intermixing and periodic self-assembly of borophene line defects. *Nat. Mater.* **17**, 783–788 (2018).

38. G. Kresse, J. Furthmüller, Efficient iterative schemes for ab initio total-energy calculations using a plane-wave basis set. *Phys. Rev. B* **54**, 11169–11186 (1996).
39. J. P. Perdew, K. Burke, M. Ernzerhof, Generalized gradient approximation made simple. *Phys. Rev. Lett.* **77**, 3865–3868 (1996).
40. P. E. Blöchl, Projector augmented-wave method. *Phys. Rev. B* **50**, 17953–17979 (1994).

**Acknowledgments**

**Funding**: H.L., A.T., C.L., and M.C.H acknowledge support from the Office of Naval Research (ONR N00014-21-1-2679) and the National Science Foundation Materials Research Science and Engineering Center (NSF DMR-2308691). Q.R. and B.I.Y acknowledge support from the Electronics Division of the Army Research Office (W911NF-16-1-0255) and the Office of Naval Research (N00014-22-1-2753).

**Author contributions:** H.L. and M.C.H. conceived the project. H.L. prepared the samples, performed STM/STS measurements, and conducted data analysis with help from A.T and C.L. Q.R. and B.I.Y. designed the models, and Q.R. performed the DFT calculations. H.L. wrote the manuscript with feedback from all authors.

**Competing interests:** The authors declare that they have no competing interests.

**Data and materials availability:** All data needed to evaluate the conclusions in the paper are present in the paper and/or the Supplementary Information. Additional data related to this paper may be requested from the authors.

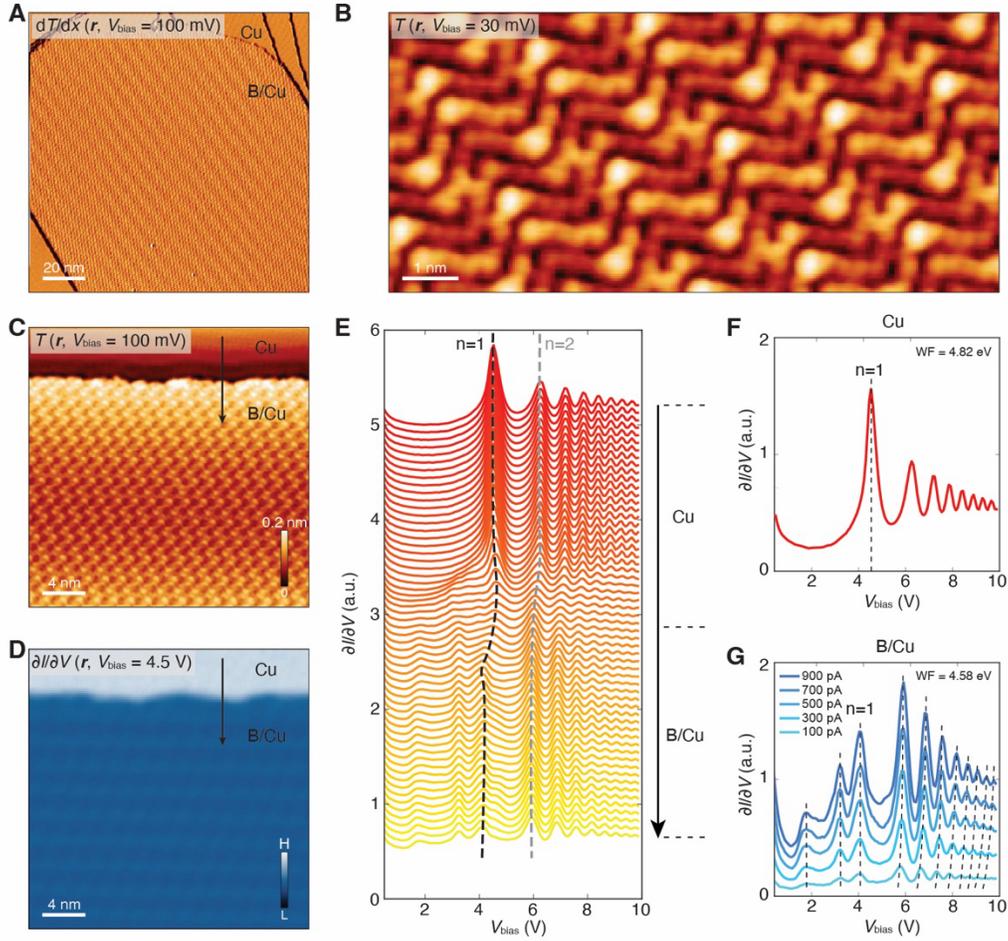

**Fig. 1. Boron deposition on Cu(111) surface and FER measurements.** (**A**) Derivative STM topography ($V_{bias}$ = 100 mV, $I_{setpoint}$ = 100 pA) of the as-grown B/Cu structure on a Cu(111) substrate. (**B**) Atomic-resolution STM image ($V_{bias}$ = 30 mV, $I_{setpoint}$ = 100 pA) of the B/Cu structure. (**C**) Zoomed-in STM topography ($V_{bias}$ = 100 mV, $I_{setpoint}$ = 300 pA) of a domain boundary between B/Cu and pristine Cu(111). (**D**) Constant-current $\partial I/\partial V$ map at $V_{bias}$ = 4.5 V extracted from FER spectroscopic mapping taken in the same area as (C). (**E**) A series of FER spectra across the boundary (black line in (C) and (D)). The purple and gray dashed lines trace the evolution of the $n$ = 1 and $n$ = 2 FER peaks, respectively. (**F**) The averaged FER spectrum on a 1 nm × 1 nm area of Cu(111). (**G**) FER spectra of B/Cu at different tip-sample distances, where the bias voltage is fixed at 100 mV and the setpoint currents vary from 100 pA to 900 pA. Each spectrum is averaged over the same 1 nm × 1 nm square on a B/Cu region.

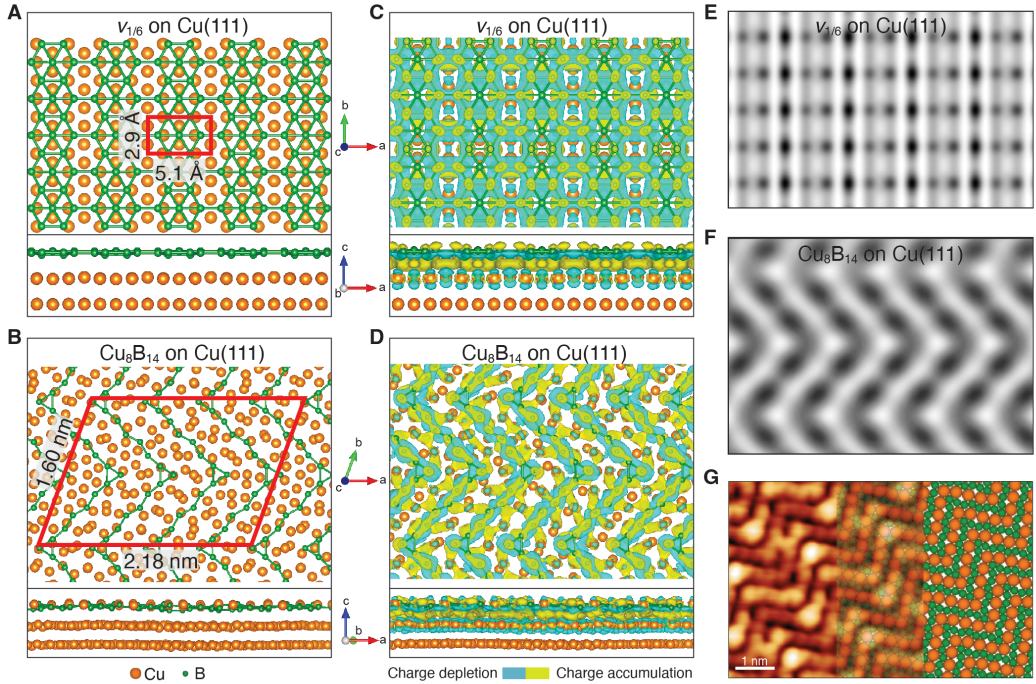

**Fig. 2. DFT calculations of borophene and copper boride on Cu(111).** (**A**) Top and side view of $v_{1/6}$ borophene on top of Cu(111). The red rectangle depicts the unit cell of $v_{1/6}$ borophene. (**B**) Top and side view of $Cu_8B_{14}$ copper boride on top of Cu(111). The red rhomboid depicts a superlattice in which the corner B atoms sit atop the second layer Cu atoms. (**C**,**D**) Calculated charge redistribution of the corresponding structures in (A) and (B), respectively. (**E**,**F**) Simulated STM images of $v_{1/6}$ borophene and $Cu_8B_{14}$ copper boride, respectively. (**G**) Experimental STM topography (cropped from Fig. 1B) overlapped with the $Cu_8B_{14}$ structural model.

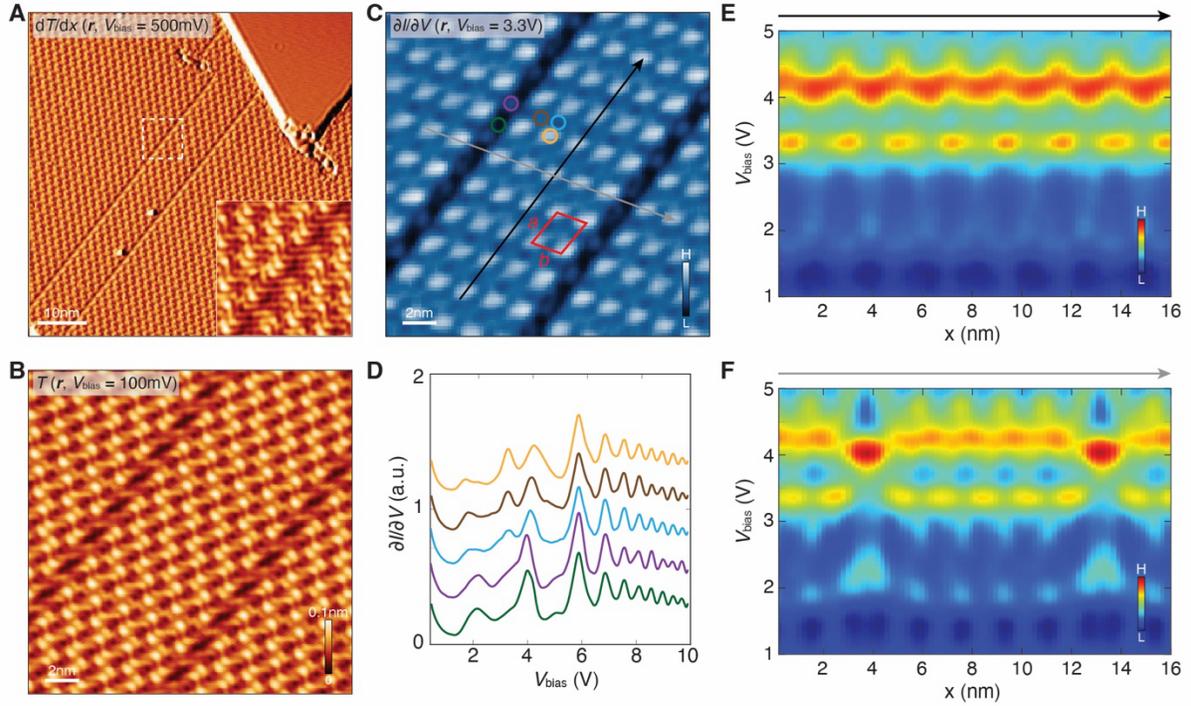

**Fig. 3. FER mapping on B/Cu with line defects.** (**A**) Derivative STM topography ($V_{bias}$ = 500 mV, $I_{setpoint}$ = 100 pA) of a B/Cu region with line defects. Inset shows the high-resolution image of the white dashed square. (**B**) STM image of the B/Cu area where the FER mapping is taken. (**C**) The $\partial I/\partial V$ map at $V_{bias}$ = 3.3 V extracted from the FER mapping. Apparent rhomboid superlattices can be resolved with lattice constants $a \sim 2.14$ nm and $b \sim 1.57$ nm. (**D**) Five typical FER spectra averaged over the colored circles in (C). The spectra are offset vertically for clarity. (**E,F**) FER spectra (1.0-5.0 V) extracted along the black and gray lines in (C), respectively.

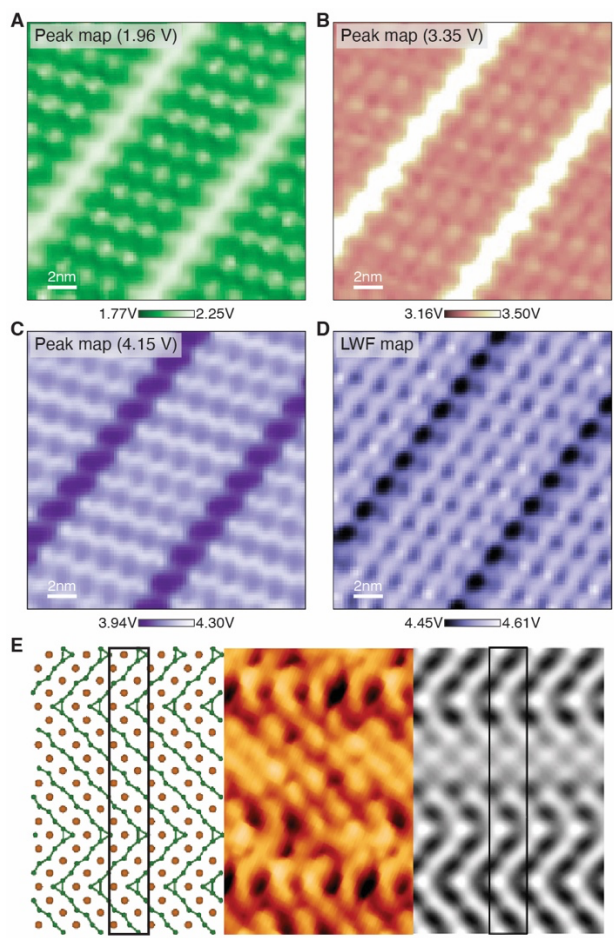

**Fig. 4. FER energy maps and cross-correlations.** (**A-C**) Peak maps near 1.96 V, 3.35 V, and 4.15 V, respectively. Gaussian fittings are used to extract the peak positions. (**D**) LWF map extracted from a fitting of Stark-shifted image potential states (peaks after $n = 2$). (**E**) From left to right: structural model, experimental STM topography, and simulated STM image of a B/Cu line defect. Black rectangles indicate the unit cell utilized in DFT calculations.

# Supplementary Materials for

## Atomic-resolution structural and spectroscopic evidence for the synthetic realization of two-dimensional copper boride

Hui Li[1], Qiyuan Ruan[2], Cataldo Lamarca[1], Albert Tsui[1], Boris I. Yakobson[2,*] and Mark C. Hersam[1,3,4,*]

* Corresponding authors: biy@rice.edu and m-hersam@northwestern.edu

**This PDF file includes:**

Figs. S1.  Height profile of Cu(111) steps after growth.
Figs. S2.  Constant-height differential conductance spectroscopy.
Figs. S3.  Constant-current $\partial I/\partial V$ spectra at negative energies.
Figs. S4.  Field-emission resonance spectroscopic mapping.
Figs. S5.  2D cross-correlation between different energy maps.

**The following video files will be available when the manuscript is published:**

Movie S1. FER mapping across a boundary between B/Cu and pristine Cu(111).
Movie S2. FER mapping over a B/Cu region with line defects.

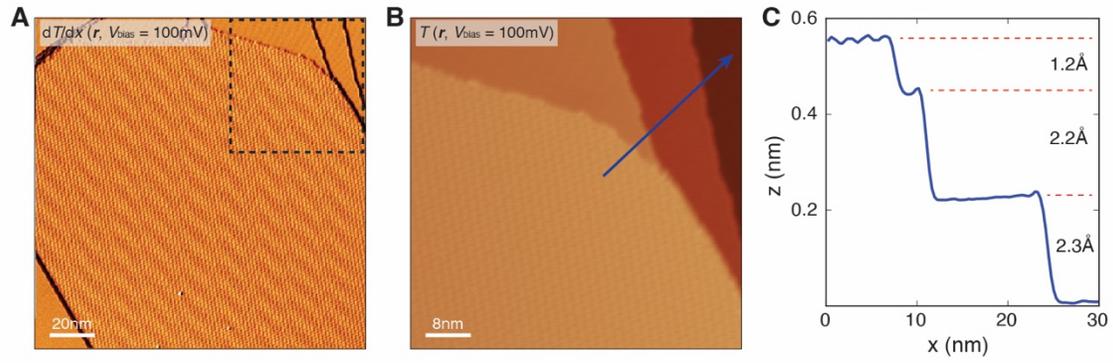

**Fig. S1. Height profile of Cu(111) steps after growth.** (**A**) Derivative STM topography ($V_{bias}$ = 100 mV, $I_{setpoint}$ = 100 pA) of a large B/Cu domain. (**B**) Zoomed-in STM image of the black dashed square in (A). Colors are re-scaled to show maximum contrast. (**C**) The height profile along the blue line in (B).

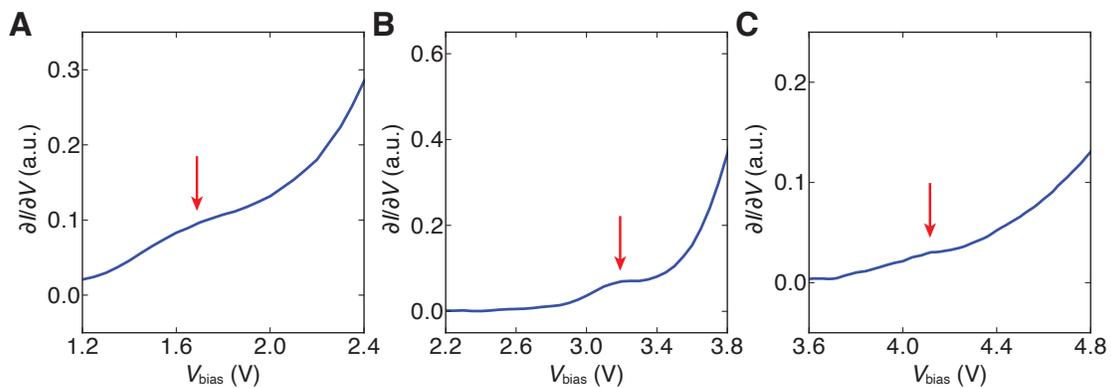

**Fig. S2. Constant-height differential conductance spectroscopy.** (A-C) Constant-height $\partial I/\partial V$ spectra taken for different energy ranges. Each spectrum is averaged over a 1 nm × 1 nm area of the B/Cu region, and the red arrows indicate the energy positions corresponding to the peaks observed in FER measurements.

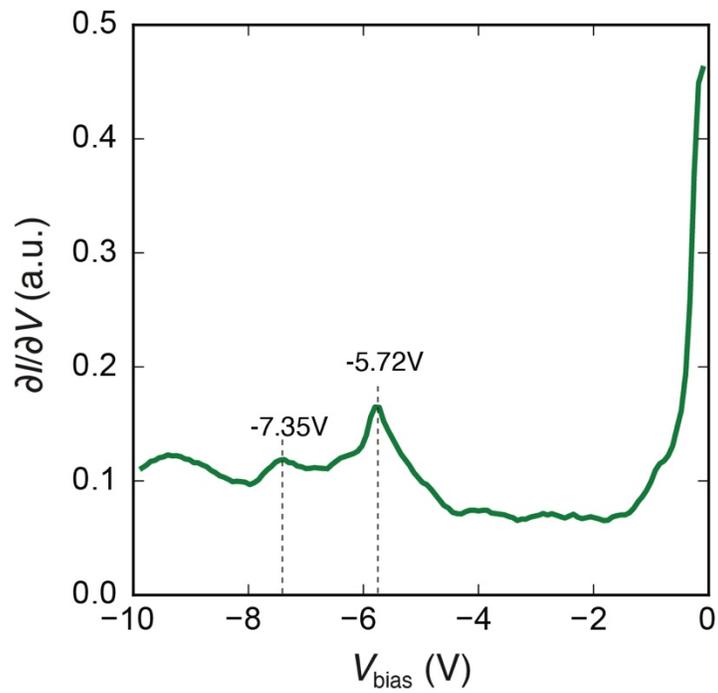

**Fig. S3. Constant-current $\partial I/\partial V$ spectra at negative energies.** $\partial I/\partial V$ measurement at negative bias voltages on the B/Cu region (averaged over a 1 nm × 1 nm area). The peaks do not reflect the resonances in this case, but instead are related to the valence bands of the structure.

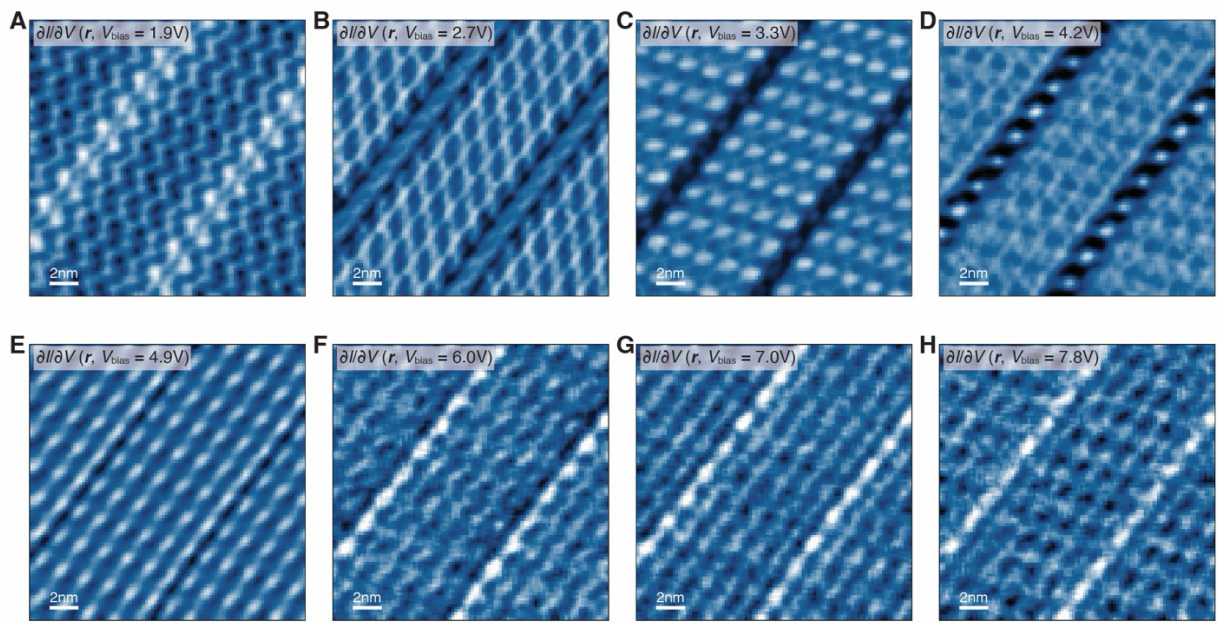

**Fig. S4. Field-emission resonance spectroscopic mapping.** (**A-H**) FER maps across B/Cu line defects at different energies.

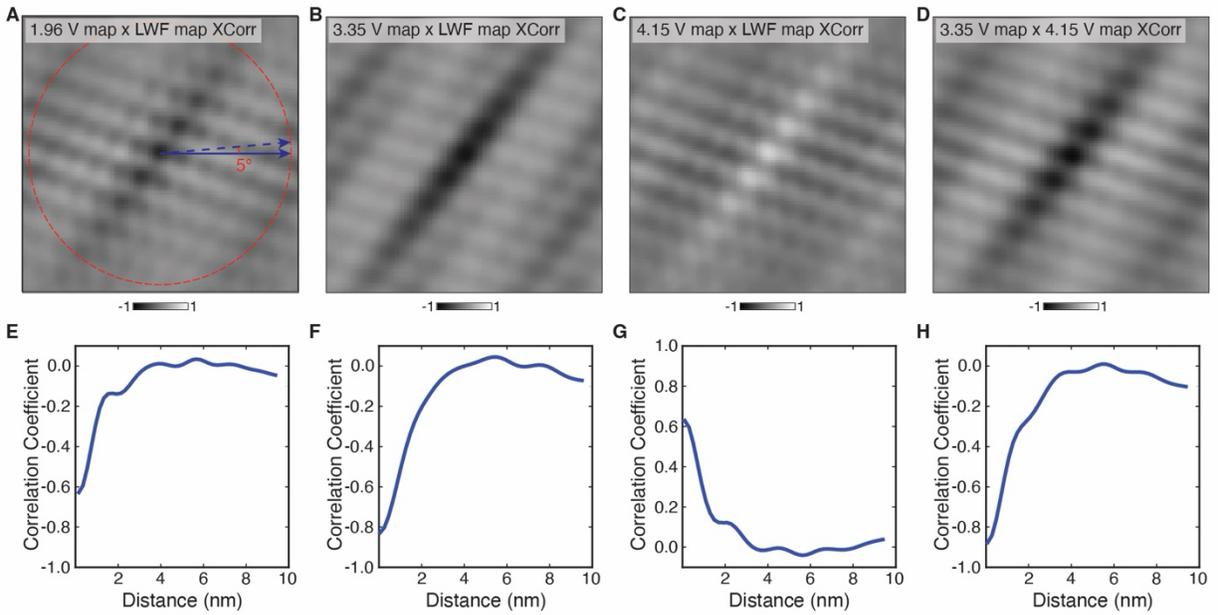

**Fig. S5. 2D cross-correlation between different energy maps.** (**A-C**) 2D cross-correlation images between the three peak maps (1.96 V, 3.35 V and 4.15 V) and the LWF map, respectively. (**D**) 2D cross-correlation image between the 3.35 V and 4.15 V peak maps. (**E-H**) The circularly averaged line profiles from the image center to edge of (A-D), respectively. Each line profile is extracted by averaging the blue lines circulating around the image center by every 5°, as shown in (A).